# Polarization-Dependent Three-Dimensional Angle-Resolved Photoemission Spectroscopy of BaFe$_{1.8}$Co$_{0.2}$As$_2$

Tetsuya HAJIRI[1,2], Takahiro ITO[1,3], Masaharu MATSUNAMI[2,4], Byeong Hun MIN[5], Yong Seung KWON[5], Shin-ichi KIMURA[2,4,*]

[1]*Graduate School of Engineering, Nagoya University, Nagoya 464-8603, Japan*
[2]*UVSOR Facility, Institute for Molecular Science, Okazaki 444-8585, Japan*
[3]*Nagoya University Synchrotron Radiation Research Center, Nagoya University, Nagoya, 464-8603, Japan*
[4]*School of Physical Sciences, The Graduate University for Advanced Studies (SOKENDAI), Okazaki 444-8585, Japan*
[5]*Department of Emerging Materials Science, DGIST, Daegu 711-873, Republic of Korea*

*E-mail: hajiri.tetsuya@f.mbox.nagoya-u.ac.jp, kimura@fbs.osaka-u.ac.jp*

We performed polarization- and photon-energy-dependent angle-resolved photoemission spectroscopy of a slightly overdoped iron pnictide superconductor, BaFe$_{1.8}$Co$_{0.2}$As$_2$, to clarify the three-dimensional electronic structure including its orbital characters at the Brillouin zone center. Two hole Fermi surfaces (FSs) with $d_{xz/yz}$ and $d_{xy/x^2-y^2}$ orbitals were observed but $d_{z^2}$ hole FS, which has nodes according to a theory of the spin-fluctuation superconductivity mechanism, did not appear. These results suggest that no node will appear at hole FSs at the zone center.

## 1. Introduction

The conduction bands as well as Fermi surfaces (FSs) of iron pnictide superconductors are very complex because of their Fe 3$d$ multi-orbital characters in spite that cuprates have only $d_{x^2-y^2}$ orbital [1]. In the iron pnictides, the superconductivity commonly appears in the vicinity of the antiferromagnetic (AFM) and/or spin-density-wave (SDW) phase [2]. Actually, NMR studies claimed that iron pnictides have strong spin fluctuation [3], and neutron scattering studies also pointed out that a magnetic scattering vector of ($\pi,\pi,\pi$) exists [4]. These results suggest that the relation between superconductivity and spin fluctuation is important. According to a theory of the spin-fluctuation superconductivity mechanism [5,6], in the case that the shape of hole FSs at the Brillouin zone center is similar to that of electron FSs, the spin

---

* present address: *Graduate School of Frontier Biosciences, Osaka University, 1-3 Yamadaoka, Suita, Osaka 565-0871, Japan*

fluctuation would be induced by the FS nesting. Then a fully gapped $s\pm$-wave superconductivity is realized.

$BaFe_{2-x}Co_xAs_2$ system is characterized by the strong spin fluctuation with an AFM wave vector between hole and electron FSs [7,8], which is similar to other iron pnictides. Hence, the origin of the high superconducting transition temperature ($T_c$) of this system has been proposed to be due to spin fluctuation as well as SDW. In the underdoped region of x ≤ 0.15, a fully opened superconducting gap has been observed [9]. In the overdoped region of x > 0.15, on the other hand, the partially closed superconducting gap with nodes on at least one FS has been suggested by using low-temperature specific heat experiments [10,11] and a heat transport measurement [12]. Therefore the superconducting property in the overdoped region is not the same as that in the underdoped region. The crossover between nodeless and nodal electronic structure of $BaFe_{2-x}Co_xAs_2$ system informs us the effective interaction of a Cooper pair formation. In particular, the position of nodes provides us important information for understanding the superconductivity mechanism of iron pnictide superconductors. One possibility of the appearance of nodes is undeveloped of the spin fluctuation. The spin fluctuation can be developed only due to intra-orbital scatterings [13]. If there is no intra-orbital scattering at a FS, the spin fluctuation does not appear, and then the nodes are induced on an FS without spin fluctuation. For instance, $BaFe_2As_{2-x}P_x$ system has $d_{z^2}$ orbital at only one FS around the Z point, therefore the spin fluctuation cannot develop at $d_{z^2}$ orbital FS and hence the node appears around there [14,15]. These results suggest the interplay between orbital characters and spin fluctuation plays an important role in the superconducting property for iron pnictides. To probe the nodes along the $k_z$ direction, it is necessary to investigate the three-dimensional (3D) electronic structure including the orbital characters.

In this paper, we report the electronic structure including the orbital characters of slightly overdoped $BaFe_{1.8}Co_{0.2}As_2$ using a polarization-dependent 3D angle-resolved photoemission (3D-ARPES) spectroscopy. We clarify that there are two hole bands at the zone center of $BaFe_{1.8}Co_{0.2}As_2$ and they show a rigid band shift from those of optimally-doped $BaFe_{2-x}Co_xAs_2$. The orbital characters of the two hole bands are determined as $d_{xz/yz}$ and $d_{xy/x^2-y^2}$ orbitals, and the $d_{z^2}$ orbital does not appear. Based on the obtained orbital characters, we discuss the possibility of the nodal superconductivity of $BaFe_{1.8}Co_{0.2}As_2$.

## 2. Experiment

The 3D-ARPES experiments were performed on single crystals of slightly overdoped $BaFe_{1.8}Co_{0.2}As_2$ with $T_c$ ~ 22.5 K. ARPES measurements were carried out at the "SAMRAI" end station of the undulator beamline 7U of UVSOR-II in the Institute for Molecular Science [16]. In this beamline, two linearly polarized lights parallel to and perpendicular to the mirror plane (namely, P and S polarizations) can be irradiated to a sample without changing the sample position. The total energy and angular resolutions were set to ~ 10 meV and ~ 0.17°, respectively. All of the measurements were performed on *in-situ* cleaved samples at $T$ = 12 K in an ultrahigh vacuum of about $8 \times 10^{-9}$ Pa. The Fermi level ($E_F$) of samples was calibrated with reference to that of evaporated gold film.

By using linearly polarized light, the orbital symmetry of electronic states can be determined using the dipole selection rule. The photoemission intensity is proportional to the dipole transition probability $<f|A\cdot p|i>$, where $A$ and $p$ are the vector potential of the electromagnetic field and the momentum operator, respectively, and $|f>$ and $|i>$ are the final- and initial-state wave functions, respectively. In the case of normal emission, since the final state $|f>$ has an even symmetry with respect to the mirror plane [17,18], the nonvanishing condition of the dipole transition $<f|A\cdot p|i>$ is that the initial state $|i>$ must have the same symmetry (even/odd) as the dipole operator $A\cdot p$. For example, if $A\cdot p$ is even (odd) corresponding to the $P$ ($S$) polarization, then the initial states with even (odd) symmetry should be reflected in the ARPES spectra. In Table I, we summarize the polarization-dependent sensitivity for the Fe $3d$ orbitals along the Γ-M direction.

**Table I.** The possibility to detect $3d$ orbitals along Γ-M high-symmetry directions.

|        | $d_{xy}$ | $d_{yz}$ | $d_{xz}$ | $d_{x^2-y^2}$ | $d_{z^2}$ |
|--------|----------|----------|----------|---------------|-----------|
| P pol. |          |          | ○        | ○             | ○         |
| S pol. | ○        | ○        |          |               |           |

## 3. Results & discussion

Since iron pnictides have 3D electronic structure [19-21], it is necessary to determine the photon energies that correspond to the high-symmetry $k_z$ points at the normal emission by using the following equation:

$$k_z = \left(\frac{2m}{\hbar^2}\right)^{1/2} [(h\nu - \Phi - E_B)\cos^2\theta + V_0], \quad (1)$$

where $m$, $\hbar$, $h\nu$, $\Phi$, $E_B$, $\theta$, $V_0$ are the free electron mass, Planck's constant, photon energy, work function, electron binding energy, the emission angle of photoelectrons from sample's surface normal direction and inner potential, respectively. Since $V_0$ depends on material, $V_0$ and $h\nu$ of the high symmetry $k_z$ points should be experimentally determined using photon-energy-dependent ARPES in the normal emission geometry.

Figures 1(a) and 1(b) show photon-energy-dependent ARPES spectra and its image, respectively, at the normal emission ($\theta = 0°$) by changing the photon energy by 1 eV from 16 to 37 eV using $S$ polarized light. The photon-energy dependence is clearly visible. The electron band at $h\nu \sim 19$ eV and the hole band at $h\nu \sim 30$ eV are observed. On the basis of Eq. (1), $V_0$ is experimentally determined as 17.5 eV, and then $h\nu = 19$ and 30 eV correspond to the Γ and Z points, respectively.

Figure 2 shows the polarization-dependent in-plane ARPES images near Γ and Z points ($h\nu = 20$ and 30 eV, respectively) along the Γ-M and Z-A directions. The solid circles in each panel indicate the band dispersions determined by the peak positions of the energy-distribution curves (EDCs) and the momentum-distribution curves (MDCs). There are two hole bands near $E_F$ at both the Γ and Z points. However, only one hole band crosses $E_F$ at the Γ point even though both two hole bands cross $E_F$ at the Z point. The Fermi wavenumber ($k_F$) of the outer hole band at the Γ point is the same as that at the Z point, indicating a two-dimensionality of the outer hole FS. On the other hand, the inner hole band has a strong three-dimensionality with a small hole FS only near the Z point.

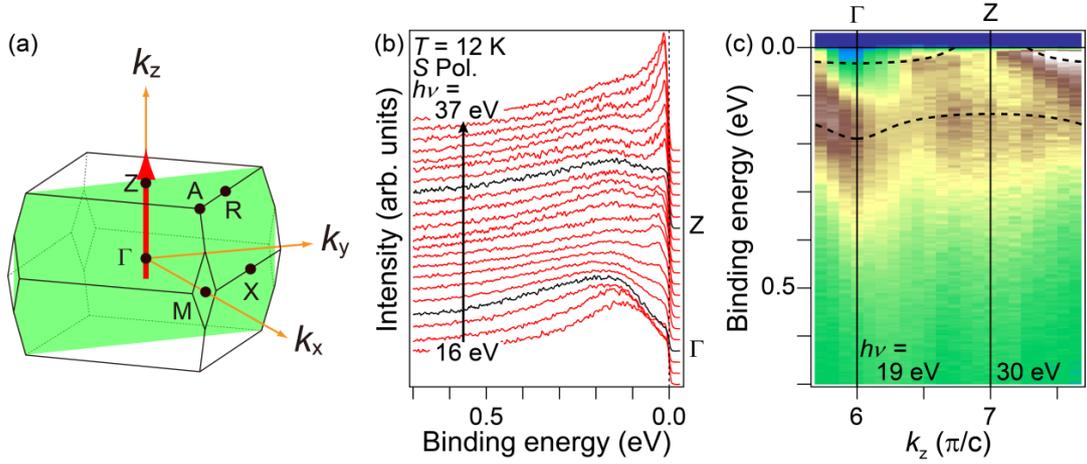

**Fig. 1.** (Color online) (a) Brillouin Zone of $BaFe_{1.8}Co_{0.2}As_2$. Photon-energy-dependent ARPES spectra (b) and its intensity plot (c) of $BaFe_{1.8}Co_{0.2}As_2$ at normal emission. The dashed curves indicate the traces of peaks.

Since the outer hole band is observed in both polarizations and shows a two-dimensionality, the orbital character can be attributed to the combination of $d_{xy}$ and $d_{x^2-y^2}$ orbitals owing to the selection rule [17, 18] as shown in Table I. On the other hand, the inner hole band is observed in both polarizations and shows a three-dimensionality. So the inner hole band can be attributed to the combination of the $d_{xz}$ and $d_{yz}$ orbitals as shown in Table I. We note that the matrix element of $d_{z^2}$ orbital varies differently on photon energy [22], but there is no prominent feature of $d_{z^2}$ orbital at several photon energies, so we conclude that there are no hole band with the $d_{z^2}$ orbital character. The orbital characters of all hole bands are consistent with those of optimally-doped $BaFe_{1.5}Co_{0.15}As_2$ [23]. The $k_\text{F}$s of observed two hole bands are about 0.6 and 1.4 Å$^{-1}$, which are smaller than those of optimally-doped $BaFe_{1.5}Co_{0.15}As_2$ [23] and are consistent with the values predicted by a rigid band model [24].

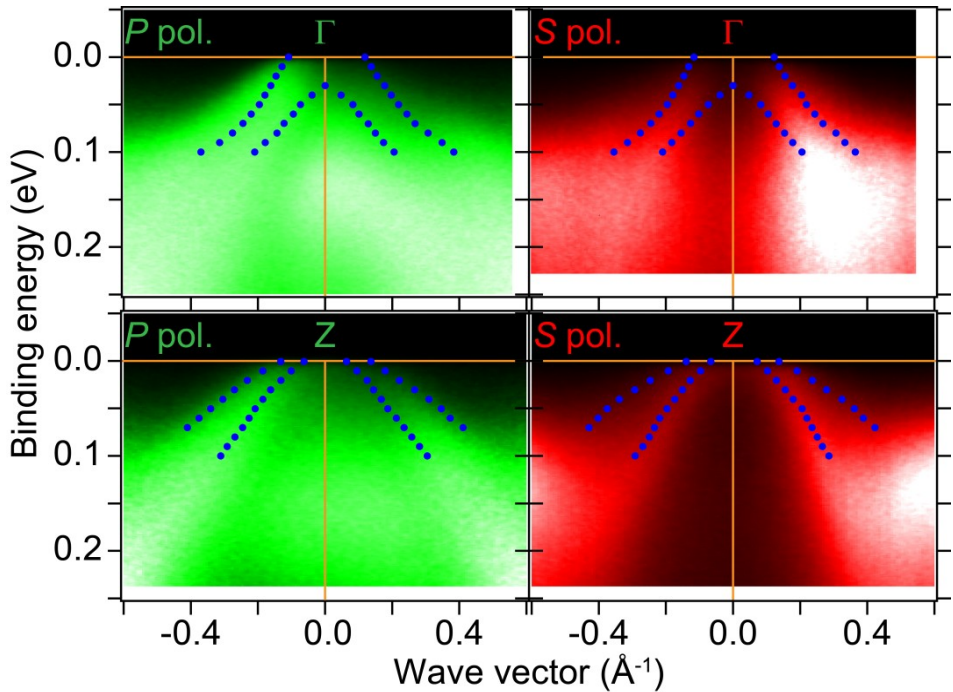

**Fig. 2.** (Color online) Polarization-dependent ARPES images of $BaFe_{1.8}Co_{0.2}As_2$ near the Γ and Z points. The left- and right-hand images were obtained using *P* and *S* polarized lights, respectively. The solid circles in each panel are the band dispersions determined by the peak positions of the EDCs and the MDCs in the range from $E_F$ to 0.1 eV.

According to the theory of the spin-fluctuation superconductivity mechanism, the superconducting-gap structure of 122-system is very sensitive to the existence of the $d_{z^2}$ orbital [14, 15]. However, here we clearly show that the $d_{z^2}$ orbital does not exist in the hole FS at the zone center of $BaFe_{1.8}Co_{0.2}As_2$. In this slightly over doped system, a node is expected to exist [10-12]. This result suggests the other possibility of the origin of nodes. One possibility is that there are nodes in the electron FSs due to an intraband scattering [25] or an orbital fluctuation [26]. The possibility should be checked in the next step.

## 4. Conclusion

We performed the polarization-dependent 3D-ARPES measurements to investigate the 3D electronic structure of hole bands of a slightly overdoped $BaFe_{1.8}Co_{0.2}As_2$. We observed the band dispersions with their orbital characters at the Γ and Z points, and confirmed there is no $d_{z^2}$ orbital on the hole FSs. These results indicate that the nodes predicted by a theory of the spin-fluctuation superconductivity mechanism don't appear in the hole FS of the slightly overdoped system. The other mechanism of an orbital fluctuation and/or an intraband scattering should be adopted.

**Acknowledgment**

The authors gratefully acknowledge M. Sakai for his technical assistance during the experiments. Part of this work was supported by the Use-of-UVSOR Facility Program (BL7U, 2011, 2012) of the Institute for Molecular Science and by a Grant-in-Aid for Scientific Research (B) from JSPS (Grant No. 22340107). The work at DGIST was partially supported by Basic Science Research Program (Grant No. 2010-0007487) and Leading Foreign Research Institute Recruitment Program (grant number 2012K1A4A3053565) through the NRF funded by MEST. T.H. was supported by Research Fellowship for Young Scientists from JSPS.